\documentclass[preprint,showpacs,preprintnumbers,amsmath,amssymb]{revtex4}

\usepackage{graphicx,color}

\begin{document}

\title{
Gravitomagnetic bending angle of light 
with finite-distance corrections 
in stationary axisymmetric spacetimes} 
\author{Toshiaki Ono}
\author{Asahi Ishihara}
\author{Hideki Asada} 
\affiliation{
Graduate School of Science and Technology, Hirosaki University,
Aomori 036-8561, Japan} 
\date{\today}

\begin{abstract} 
By using the Gauss-Bonnet theorem, the bending angle of light 
in a static, spherically symmetric and asymptotically flat spacetime 
has been recently discussed, 
especially by taking account of the finite distance from a lens object 
to a light source and a receiver 
[Ishihara, Suzuki, Ono, Asada, Phys. Rev. D {\bf 95}, 044017 (2017)]. 
We discuss a possible extension of the method of calculating 
the bending angle of light to stationary, axisymmetric 
and asymptotically flat spacetimes. 
For this purpose, we consider the light rays on the equatorial plane 
in the axisymmetric spacetime. 
We introduce a spatial metric to define the bending angle of light 
in the finite-distance situation. 
We show that the proposed bending angle of light is coordinate-invariant 
by using the Gauss-Bonnet theorem. 
The non-vanishing geodesic curvature of the photon orbit 
with the spatial metric is caused in gravitomagnetism, 
even though 
the light ray in the four-dimensional spacetime follows the null geodesic. 
Finally, we consider Kerr spacetime as an example 
in order to examine how the bending angle of light is computed 
by the present method. 
The finite-distance correction to the gravitomagnetic deflection angle 
due to the Sun's spin 
is around a pico-arcsecond level. 
The finite-distance corrections for Sgr A$^{\ast}$ also 
are estimated to be very small. 
Therefore, the gravitomagnetic finite-distance corrections for these objects 
are unlikely to be observed with present technology. 
\end{abstract}


\maketitle

\section{Introduction}
Since the experimental confirmation of the theory of general relativity 
\cite{GR} succeeded in 1919 \cite{Eddington}, 
a lot of calculations of the gravitational bending of light 
have been done not only for black holes 
\cite{Hagihara, Ch, MTW, Darwin, Bozza, Iyer, Bozza+, Frittelli, VE2000, Virbhadra, VNC, VE2002, VK2008, Zschocke}
but also for other objects such as wormholes and gravitational monopoles 
\cite{ERT, Perlick, Abe, Toki, Nakajima, Gibbons, DA, Kitamura,Tsukamoto,Izumi,Kitamura2014,Nakajima2014,Tsukamoto2014}

Gibbons and Werner (2008) proposed an alternative way of deriving 
the deflection angle of light \cite{GW2008}. 
They assumed that the source and receiver are located 
at an asymptotic Minkowskian region and they used the Gauss-Bonnet theorem 
to a spatial domain described by the optical metric, 
for which a light ray is described as a spatial curve. 
Ishihara et al. have recently extended Gibbons and Werner's idea 
in order to investigate 
finite-distance corrections 
in the small deflection case 
(corresponding to a large impact parameter case)
\cite{Ishihara2016} 
and also in the strong deflection limit 
for which the photon orbits may have the winding number 
larger than unity \cite{Ishihara2017}. 
In particular, the asymptotic receiver and source 
have not been assumed. 

However, the earlier treatments \cite{Ishihara2016, Ishihara2017} 
are limited within the spherical symmetry. 
It is not clear whether the Gauss-Bonnet method with using the optical metric 
can be extended to axisymmetric cases or not.  
This is mostly because there can exist 
off-diagonal (time-space) components of the spacetime metric 
in an axisymmetric spacetime. 
The time-space components 
seem to make it unclear whether the optical metric can be constructed. 
After the gravitational lensing by a spinning object 
\cite{ES, Ibanez, AK} 
and that by a relativistic binary \cite{Kopeikin} 
were discussed extensively by perturbative approaches such as 
the post-Newtonian approximation, 
Werner (2012) \cite{Werner2012} proposed the use of the 
Kerr-Randers optical geometry on this issue 
\cite{Kerr-bending}. 
To be more precise, he used the osculating Riemann approach 
in Finsler geometry 
in order to discuss the lensing by the Kerr black hole, 
for which the metric can be written in the Randers form. 
However, this approach requires that 
the endpoints (namely, the source and the receiver) 
of the photon orbit are in Euclidean space, 
for which angles can be easily defined. 
This requirement is mainly because  
jump angles at the vertices in the Gauss-Bonnet theorem 
are problematic in the Finsler geometry. 
Namely, it is unlikely that the Finsler geometry 
can be used for computing the finite-distance corrections. 

Therefore, the main purpose of the present paper is 
to extend the earlier formulation in Refs. \cite{Ishihara2016, Ishihara2017}, 
especially in order to examine finite-distance corrections to 
the deflection angle of light in the axisymmetric spacetime, 
for which the gravitational deflection of light 
may include gravitomagnetic effects 
(e.g. \cite{ES, Ibanez, AK, Kopeikin}). 
The geometrical setups in the present paper are not 
those in the optical geometry, 
in the sense that the photon orbit has a non-vanishing 
geodesic curvature, 
though the light ray in the four-dimensional spacetime 
obeys a null geodesic.

This paper is organized as follows. 
Section II discusses a possible extension 
to stationary and axisymmetric spacetimes. 
In particular, it is shown that 
the proposed definition of the deflection angle 
is coordinate-invariant by using the Gauss-Bonnet theorem. 
Section III uses the Kerr metric as a known example of 
the stationary and axisymmetric spacetimes 
in order to discuss how to compute 
the gravitational deflection angle of light 
by the proposed method. 
Section IV is devoted to conclusion. 
In Appendix A, the deflection angle of light is computed  
at the second order of the mass and the spin parameter 
in order to examine whether 
the deflection angle is in agreement with the known one. 
Throughout this paper, we use the unit of $G=c=1$, 
and the observer may be called the receiver 
in order to avoid a confusion between $r_O$ and $r_0$ by using $r_R$.

\section{Extension to axisymmetric spacetimes}
\begin{figure}
\includegraphics[width=12cm]{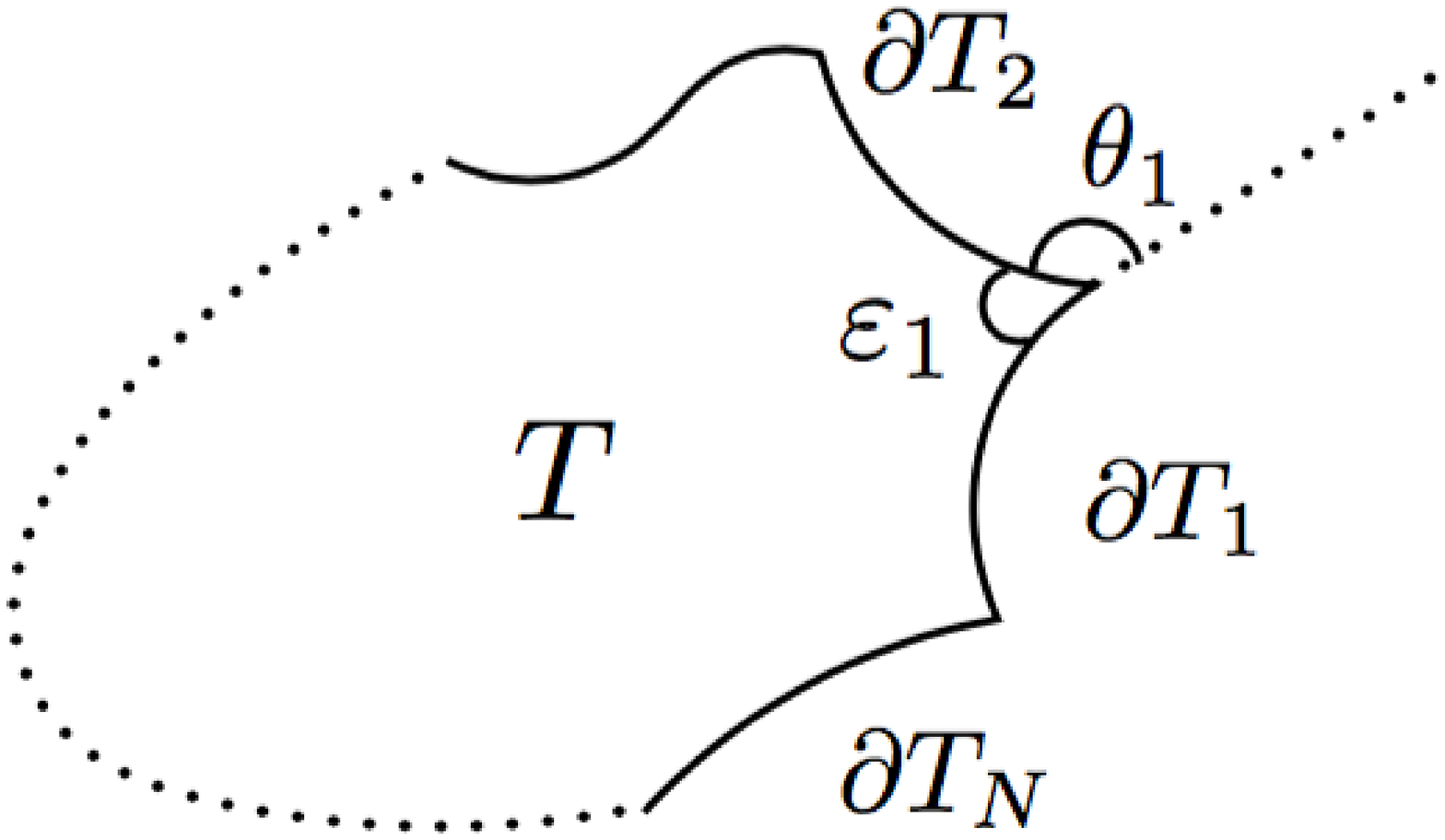}
\caption{
Schematic figure for the Gauss-Bonnet theorem. 
}
\label{fig-GB}
\end{figure}

Henceforth, we assume a stationary and axisymmetric spacetime, 
for which we shall define the gravitational deflection angle of light 
by using the Gauss-Bonnet theorem: 
Suppose that 
$T$ is a two-dimensional orientable surface with boundaries $\partial T_a$ 
($a=1, 2, \cdots, N$) that 
are differentiable curves. 
See Figure \ref{fig-GB}. 
Let the jump angles between the curves be $\theta_a$ 
($a=1, 2, \cdots, N$). 
Then, the Gauss-Bonnet theorem can be expressed as 
\cite{GB-theorem}
\begin{eqnarray}
\iint_{T} K dS + \sum_{a=1}^N \int_{\partial T_a} \kappa_g d\ell + 
\sum_{a=1}^N \theta_a = 2\pi , 
\label{localGB}
\end{eqnarray}
where 
$K$ denotes the Gaussian curvature of 
the surface $T$, 
$dS$ is the area element of the surface, 
$\kappa_g$ means the geodesic curvature of $\partial T_a$, 
and $\ell$ is the line element along the boundary. 
The sign of the line element is chosen such that it is 
compatible with the orientation of the surface.

\subsection{Stationary, axisymmetric spacetime}
We consider a stationary axisymmetric spacetime. 
The line element for this spacetime is \cite{Lewis,LR,Papapetrou}
\begin{align}
ds^2=&g_{\mu\nu}dx^{\mu}dx^{\nu} \notag\\
=&-A(y^p,y^q)dt^2-2H(y^p,y^q)dtd\phi \notag\\
&+F(y^p,y^q)(\gamma_{pq}dy^pdy^q)+D(y^p,y^q)d\phi^2 , 
\label{ds2-general}
\end{align}
where $\mu, \nu$ run from $0$ to $3$, 
$p, q$ take $1$ and $2$, 
$t$ and $\phi$ coordinates are associated with the Killing vectors,  
and $\gamma_{pq}$ is a two-dimensional symmetric tensor. 
It is more convenient to 
reexpress this metric into 
a form in which $\gamma_{pq}$ is diagonalized. 
The present paper prefers the polar coordinates 
rather than the cylindrical ones, 
because 
the Kerr metric in the polar coordinates is considered in Section III. 
In the polar coordinates, 
Eq. (\ref{ds2-general}) becomes \cite{Metric}
\begin{align}
ds^2=&-A(r,\theta)dt^2-2H(r,\theta)dtd\phi \notag\\
&+B(r,\theta)dr^2+C(r,\theta)d\theta^2+D(r,\theta)d\phi^2 . 
\label{ds2-axial}
\end{align}

The null condition $ds^2 = 0$ is solved for $dt$ as \cite{AK}
\begin{align}
dt=& \sqrt{\gamma_{ij} dx^i dx^j} +\beta_i dx^i , 
\label{opt} 
\end{align}
where 
$i, j$ run from $1$ to $3$, 
$\gamma_{ij}$ and $\beta_i$ are defined as 
\begin{align}
\gamma_{ij}dx^idx^j \equiv&
\frac{B(r,\theta)}{A(r,\theta)}dr^2
+\frac{C(r,\theta)}{A(r,\theta)}d\theta^2
+\frac{A(r,\theta)D(r,\theta)+H^2(r,\theta)}{A^2(r,\theta)}d\phi^2 , 
\label{gamma}
\\
\beta_idx^i \equiv& -\frac{H(r,\theta)}{A(r,\theta)} d\phi . 
\label{beta}
\end{align}

This spatial metric $\gamma_{ij} (\neq g_{ij})$ 
may define the arc length ($\ell$) along the light ray as 
\begin{align}
d\ell^2 \equiv \gamma_{ij} dx^i dx^j , 
\end{align}
for which 
$\gamma^{ij}$ is defined by 
$\gamma^{ij}\gamma_{jk} = \delta^i_{~k}$. 
Note that $\ell$ defined in this way is an affine parameter 
along the light ray. 
See e.g. Appendix of Ref. \cite{AK} for the proof on the affine parameter 
\cite{comment-gamma}.

$\gamma_{ij}$ defines a 3-dimensional Riemannian space ${}^{(3)}M$ 
in which 
the motion of the photon is described as a motion in a spatial curve. 
The unit tangential vector along the spatial curve 
is defined as 
\begin{align}
e^i \equiv \frac{dx^i}{d\ell} . 
\end{align}

The light ray  
follows the Fermat's principle \cite{Perlick}. 
By using the variational principle, 
this gives the equation for the light ray as \cite{AK} 
\begin{align}
e^i_{~|k}e^k=a^i , 
\label{EL}  
\end{align}
where 
$|$ denotes the covariant derivative with $\gamma_{ij}$ 
and 
$a^i$ is defined as 
\begin{align}
a^i \equiv 
\gamma^{ij}(\beta_{k|j}-\beta_{j|k})e^k . 
\label{ai}
\end{align}
Here, 
\begin{align}
e^i_{~|k} e^k = \frac{de^i}{d\ell}+{}^{(3)}\Gamma^i_{~jk}e^je^k , 
\end{align}
where 
${}^{(3)}\Gamma^i_{~jk}$ denotes the Christoffel symbol 
associated with $\gamma_{ij}$. 

The vector $a^i$ is the spatial vector that means the acceleration 
originated from $\beta_i$. 
In particular, $a^i$ is caused in gravitomagnetism 
as discussed below in more detail. 
This has an analogy as the acceleration by the Lorentz force 
$\propto \vec{v} \times (\vec{\nabla} \times \vec{A}_m$) 
in electromagnetism, 
where $\vec{A}_m$ denotes the magnetic vector potential. 

We should note that $\gamma_{ij}$ is not an induced metric. 
As a result, the photon orbit 
can deviate from a geodesic in ${}^{(3)}M$ with $\gamma_{ij}$ 
if $\beta_i \neq 0$, 
even though the light ray in the four-dimensional spacetime 
follows the null geodesic.

For a stationary and spherically symmetric spacetime, 
one can always find a set of suitable coordinates 
such that $g_{0i}$ can vanish to lead to $a^i = 0$. 
In this case, the photon orbit 
becomes a spatial geodesic curve in ${}^{(3)}M$. 

The present paper discusses an extension to axisymmetric 
cases, which allow $g_{0i} \neq 0$. 
Therefore, we have to take account of non-zero $\kappa_g$ 
along the photon orbit in the Gauss-Bonnet theorem. 
This non-vanishing $\kappa_g$ of the photon orbit 
makes a crucial difference from the previous papers 
\cite{Ishihara2016,Ishihara2017}

\subsection{Geodesic curvature and equatorial plane}
Let us imagine a parameterized curve in a surface. 
The geodesic curvature of the parameterized curve 
is the surface-tangential component 
of acceleration (namely curvature) of the curve, 
while the normal curvature is the surface-normal component. 
The normal curvature has nothing to do with the present paper. 
The geodesic curvature can be defined in the vector form as 
(e.g. \cite{Math})
\begin{align}
\kappa_g \equiv \vec{T}^{\prime} \cdot \left(\vec{T} \times \vec{N}\right) , 
\label{kappag-vector}
\end{align}
where we assume a parameterized curve with a parameter, 
$\vec{T}$ is the unit tangent vector for the curve 
by reparameterizing the curve using its arc length, 
$\vec{T}^{\prime}$ is its derivative with respect to the parameter, 
and $\vec{N}$ is the unit normal vector for the surface. 
In this paper, 
Eq. (\ref{kappag-vector}) can be rewritten in the tensor form as 
\begin{align}
\kappa_g = \epsilon_{ijk} N^i a^j e^k , 
\label{kappag-tensor}
\end{align}
where $\vec{T}$ and $\vec{T}^{\prime}$ correspond to 
$e^k$ and $a^j$, respectively. 
Here, the Levi-Civita tensor 
$\epsilon_{ijk}$ is defined by 
$\epsilon_{ijk} \equiv \sqrt{\gamma}\varepsilon_{ijk}$, 
where 
$\gamma \equiv \det{(\gamma_{ij})}$, 
and $\varepsilon_{ijk}$ is the Levi-Civita symbol 
($\varepsilon_{123} = 1$). 
In the present paper, the space is ${}^{(3)}M$. 
Therefore, we use $\gamma_{ij}$ in the above definitions 
but not $g_{ij}$. 

For a case of $a^i \neq 0$ due to $g_{0i}$, 
there can exist a non-vanishing integral 
of the geodesic curvature along the light ray 
in the Gauss-Bonnet theorem by Eq. (\ref{localGB}). 

By substituting Eq. (\ref{ai}) into $a^i$ in Eq. (\ref{kappag-tensor}), 
we obtain 
\begin{align}
\kappa_g = - \epsilon^{ijk} N_i \beta_{j|k} , 
\label{kappag-tensor2} 
\end{align}
where we use $\gamma_{ij}e^ie^j = 1$. 

Up to this point, the surface in ${}^{(3)}M$ is not specified. 
Henceforth, we focus on the equatorial motion of the photon. 
We choose $\theta = \pi/2$ as the equatorial plane. 
Then, the unit normal vector 
for the equatorial plane can be expressed as 
\begin{align}
N_p = \frac{1}{\sqrt{\gamma^{\theta\theta}}} \delta_p^{\theta} , 
\label{N}
\end{align}
where we choose the upward direction without loss of generality. 

For the equatorial case, one can show 
\begin{align}
\epsilon^{\theta p q} \beta_{q|p} 
&=-\frac{1}{\sqrt{\gamma}}\beta_{\phi,r} , 
\label{rot-beta}
\end{align}
where the comma denotes the partial derivative, 
we use $\epsilon^{\theta r \phi} = - 1/\sqrt{\gamma}$ 
and 
we note $\beta_{r,\phi} = 0$ owing to the axisymmetry. 
By using Eqs. (\ref{N}) and (\ref{rot-beta}), 
an explicit form of $\kappa_g$ in 
Eq. (\ref{kappag-tensor2}) is obtained as 
\begin{align}
\kappa_g=-\frac{1}{\sqrt{\gamma\gamma^{\theta\theta}}} \beta_{\phi,r} . 
\label{kappag-final}
\end{align}

\subsection{Impact parameter and the photon directions 
at the receiver and source} 
We study the orbit equation on the equatorial plane 
with Eq. (\ref{ds2-axial}). 
Associated with the two Killing vectors, 
there are the two constants of motion as 
\begin{align}
E&=A(r)\dot{t}+H(r)\dot{\phi} , 
\label{EEE} 
\\
L&=D(r)\dot{\phi}-H(r)\dot{t} , 
\label{L}
\end{align}
where the dot denotes the derivative with respect to the affine parameter. 

As usual, we define the impact parameter as 
\begin{align}
b &\equiv \frac{L}{E} 
\notag\\ 
&=\frac{-H(r)\dot{t}+D(r)\dot{\phi}}{A(r)\dot{t}+H(r)\dot{\phi}} 
\notag\\
&=\cfrac{-H(r)+D(r)\cfrac{d\phi}{dt}}{A(r)+H(r)\cfrac{d\phi}{dt}} . 
\label{b}
\end{align}

In terms of the impact parameter $b$, 
$ds^2=0$ leads to the orbit equation on the equatorial plane as 
\begin{align}
\left(\frac{dr}{d\phi}\right)^2
=\frac{A(r)D(r)+H^2(r)}{B(r)}
\frac{D(r)-2H(r)b-A(r)b^2}{\left[H(r)+A(r)b\right]^2} , 
\label{OE}
\end{align}
where we use Eq. (\ref{ds2-axial}). 
Let us introduce $u \equiv 1/r$ 
to rewrite the orbit equation as 
\begin{align}
\left(\frac{du}{d\phi}\right)^2
= F(u) , 
\label{OE-2}
\end{align}
where $F(u)$ is 
\begin{align}
F(u) 
=\frac{u^4 (AD+H^2) (D-2Hb-Ab^2)}{B (H+Ab)^2} . 
\label{F-axial}
\end{align}

Finally, we examine the angles at the receiver and source positions. 
The unit tangent vector along the photon orbit in ${}^{(3)}M$ 
is $e^i$. 
On the equatorial plane, its components are obtained as 
\begin{align}
e^i=\frac{1}{\xi} \Big(\frac{dr}{d\phi}, 0, 1 \Big) . 
\label{ei}
\end{align}
Here, $\xi$ satisfies 
\begin{align}
\frac{1}{\xi}=\frac{A(r)[H(r)+A(r)b]}{A(r)D(r)+H^2(r)} , 
\label{xi}
\end{align}
which can be derived from $\gamma_{ij} e^i e^j = 1$ 
by using Eq. (\ref{OE}).

The unit radial vector in the equatorial plane is 
\begin{align}
R^i= \Big(\frac{1}{\sqrt{\gamma_{rr}}}, 0, 0 \Big) , 
\label{R}
\end{align}
where we choose the outgoing direction for a sign convention.

Therefore, we can define the angle measured from 
the outgoing radial direction by 
\begin{align}
\cos \Psi \equiv& 
\gamma_{ij} e^i R^j 
\notag\\
=& \sqrt{\gamma_{rr}}
\frac{A(r)[H(r)+A(r)b]}{A(r)D(r)+H^2(r)}
\frac{dr}{d\phi} , 
\label{cos}
\end{align}
where Eqs. (\ref{ei}), (\ref{xi}) and (\ref{R}) are used. 
This can be rewritten as 
\begin{align}
\sin\Psi 
=&\frac{H(r)+A(r)b}
{\sqrt{A(r)D(r)+H^2(r)}} , 
\label{sin}
\end{align}
where we use Eq. (\ref{OE}). 
Note that $\sin\Psi$ by Eq. (\ref{sin}) 
is more convenient in practical calculations, 
because it needs only the local quantities, 
whereas $\cos\Psi$ by Eq. (\ref{cos}) needs the derivative as $dr/d\phi$.

\subsection{Deflection angle of light}
For the equatorial case in the axisymmetric spacetime, 
we define 
\begin{equation}
\alpha \equiv \Psi_R - \Psi_S + \phi_{RS} . 
\label{alpha-axial}
\end{equation} 
This definition seems to rely on a choice 
of the angular coordinate $\phi$.

\begin{figure}
\includegraphics[width=10cm]{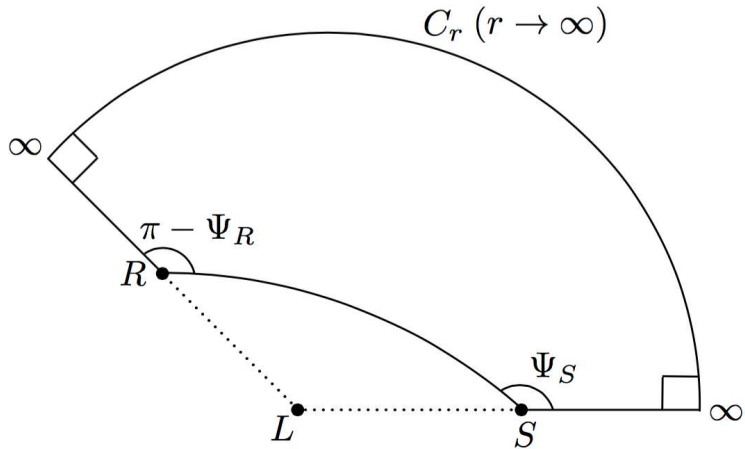}
\caption{
Quadrilateral ${}^{\infty}_{R}\Box^{\infty}_{S}$
embedded in a curved space. 
Note that the inner angle at the vertex $R$ is $\pi - \Psi_R$. 
}
\label{fig-Box}
\end{figure}

Let us consider a quadrilateral 
${}^{\infty}_{R}\Box^{\infty}_{S}$, 
which consists of the spatial curve for the light ray, 
two outgoing radial lines from R and from S 
and a circular arc segment $C_r$ 
of coordinate radius $r_C$ ($r_C \to \infty$) 
centered at the lens 
which intersects the radial lines  
through the receiver or the source. 
See Figure \ref{fig-Box} for the configuration such as 
the domain ${}^{\infty}_{R}\Box^{\infty}_{S}$. 
See also Ref. \cite{Ishihara2017} for the case that 
the winding number is larger than unity. 
For the asymptotically flat spacetime,  
$\kappa_g \to 1/r_C$ and $d\ell \to r_C d\phi$ 
as $r_C \to \infty$ (See e.g. \cite{GW2008}). 
Hence, 
$\int_{C_r} \kappa_g d\ell \to \phi_{RS}$.

By using the Gauss-Bonnet theorem Eq. (\ref{localGB}), 
Eq. (\ref{alpha-axial}) 
is rewritten as 
\begin{align}
\alpha 
=-\iint_{{}^{\infty}_{R}\square^{\infty}_{S}} K dS 
+ \int_{S}^{R} 
\kappa_g d\ell , 
\label{GB-axial}
\end{align}
where $d\ell$ is positive 
for the prograde motion of the photon 
and it is negative for the retrograde motion. 
Eq. (\ref{GB-axial}) shows that $\alpha$ is coordinate-invariant 
also for the axisymmetric case.

Up to this point, 
equations for gravitational fields are not specified. 
Therefore, the above discussion and results are not 
limited within the theory of general relativity (GR) 
but they are applicable to a certain class of modified gravity theories 
if the light ray in the four-dimensional spacetime 
obeys the null geodesic.

\section{Application to the Kerr lens} 
\subsection{Kerr spacetime and $\gamma_{ij}$} 
This section focuses on the Kerr spacetime as 
one of the most known examples with axisymmetry. 
The Boyer-Lindquist form of the Kerr metric is 
\begin{align}
ds^2=&-\left(1-\frac{2Mr}{\Sigma}\right)dt^2
-\frac{4aMr\sin^2\theta}{\Sigma}dtd\phi \notag \\
&+\frac{\Sigma}{\Delta}dr^2+\Sigma d\theta^2
+\left(r^2+a^2+\frac{2a^2Mr\sin^2\theta}{\Sigma}\right)\sin^2\theta d\phi^2 , 
\label{Kerr}
\end{align}
where $\Sigma$ and $\Delta$ are denoted as 
\begin{align}
\Sigma&\equiv r^2+a^2\cos^2\theta , 
\label{Sigma}
\\
\Delta&\equiv r^2-2Mr+a^2 . 
\end{align}

By using Eqs. (\ref{gamma}) and (\ref{beta}),  
one can see that 
$\gamma_{ij}$ and $\beta_i$ for the Kerr metric are given by 
\begin{align}
\gamma_{ij}dx^idx^j=&
\frac{\Sigma^2}{\Delta(\Sigma-2Mr)}dr^2 
+ \frac{\Sigma^2}{(\Sigma-2Mr)} d\theta^2
\notag\\
&+ \left(r^2+a^2+\frac{2a^2Mr\sin^2\theta}{(\Sigma-2Mr)}\right) 
\frac{\Sigma\sin^2\theta}{(\Sigma-2Mr)} d\phi^2 ,
\label{gamma-Kerr}
\\
\beta_idx^i=&- \frac{2aMr\sin^2\theta}{(\Sigma-2Mr)}d\phi .
\label{beta-Kerr}
\end{align}

Note that $\gamma_{ij}$ has no terms linear in the Kerr parameter $a$, 
because $g_{0i} \propto H$ enters $\gamma_{ij}$ 
in a quadratic form through $g_{0i}g_{0j} \propto H^2$ 
as shown by Eq. (\ref{gamma}). 

In order to see what is $\kappa_g$ 
for the present case, 
we employ the weak field and slow rotation approximations, 
for which $M$ and $a$ can be used as book-keeping parameters.

\subsection{Path integral of $\kappa_g$}
By substituting $\beta_i$ by Eq. (\ref{beta-Kerr}) 
into Eq. (\ref{kappag-final}), we obtain  
\begin{align}
\kappa_g
=&
-\frac{2aM}{r^2(r-2M)}
\left(
\cfrac{1-\cfrac{2M}{r}+\cfrac{a^2}{r^2}}
{1+\cfrac{a^2}{r^2}+\cfrac{2a^2M}{r^3}}
\right)^{1/2}
 \notag\\
=&-\frac{2aM}{r^3} + 
O\left(\frac{aM^2}{r^4}\right) , 
\label{kappa-Kerr}
\end{align}
where we use the weak field and slow rotation approximations 
in the last line 
and the terms of $a^nM$ $(n \geq 2)$ vanish.

The path integral of $\kappa_g$ is computed as 
\begin{align}
\int^{R}_{S}\kappa_gd\ell
=&
-\int^{R}_{S}
\left[
\frac{2aM}{r^3}
+ O\left(\frac{aM^2}{r^4}\right)  
\right]
d\ell 
\notag\\
=&
- \frac{2aM}{b^2}\int^{\phi_R}_{\phi_S} \cos\vartheta d\vartheta 
+ 
O\left(\frac{aM^2}{b^3}\right) 
\notag\\
=&
- \frac{2aM}{b^2}[\sqrt{1-b^2{u_R}^2}+\sqrt{1-b^2{u_S}^2}] 
+ O\left(\frac{aM^2}{b^3}\right) 
, 
\label{int-kappag}
\end{align}
where we assume the prograde case $d\ell > 0$ 
that the orbital angular momentum of the photon 
is aligned with the spin of the black hole 
and 
we use a linear approximation of the photon orbit 
as $r= b/\cos\vartheta + O(M, a)$ 
and $\ell = b \tan\vartheta + O(M, a)$ 
in the second line. 
Note that, in the retrograde case, the sign of $d\ell$ is negative 
and thus the magnitude of the above path integral 
remains the same but the sign is opposite.

\subsection{$\phi_{RS}$ part}
The integral of the angular coordinate $\phi$ 
becomes 
\begin{align}
\phi_{RS} 
=& 
\int^R_S d\phi 
\notag\\
=&
2\int^{u_0}_{0}\frac{1}{\sqrt{F(u)}}du 
+\int^{0}_{u_S}\frac{1}{\sqrt{F(u)}}du +\int^{0}_{u_R}\frac{1}{\sqrt{F(u)}}du , 
\label{phi-Kerr}
\end{align}
where we use the orbit equation given by Eq. (\ref{OE-2}), 
By substituting Eq. (\ref{F-axial}) into $F(u)$ in Eq. (\ref{phi-Kerr}), 
we obtain
\begin{align}
\phi_{RS} 
=&
\int^{u_0}_{u_S}\left(\frac{1}{\sqrt{{u_0}^2-u^2}}
+M\frac{{u_0}^3-u^3}{({u_0}^2-u^2)^{3/2}}
-2aM\frac{{u_0}^3(u_0-u)}{({u_0}^2-u^2)^{3/2}}\right) du 
\notag\\
&
+\int^{u_0}_{u_R}\left(\frac{1}{\sqrt{{u_0}^2-u^2}}
+M\frac{{u_0}^3-u^3}{({u_0}^2-u^2)^{3/2}}
-2aM\frac{{u_0}^3(u_0-u)}{({u_0}^2-u^2)^{3/2}}\right) du 
\notag\\ 
&+ O(M^2, a^2) 
\notag\\
=&\left(\frac{\pi}{2}-\arcsin\Big(\frac{u_S}{u_0}\Big)
+M\frac{(2u_0+u_S)\sqrt{{u_0}^2-
u_S^2}}
{u_0+u_S}
-2aM\frac{{u_0}^3\sqrt{{u_0}^2-
u_S^2}}
{{u_0}^2+u_0u_S}\right) \notag\\
&+\left(\frac{\pi}{2}-\arcsin\Big(\frac{u_R}{u_0}\Big)
+M\frac{(2u_0+u_R)\sqrt{{u_0}^2-{u_R}^2}}{u_0+u_R}
-2aM\frac{{u_0}^3\sqrt{{u_0}^2-{u_R}^2}}{{u_0}^2+u_0u_R}\right) 
\notag\\
&+ O\left(M^2 u_0^2, a^2 u_0^2\right) , 
\label{phi-Kerr1}
\end{align}
where we assume the prograde case. 
For the retrograde case, the sign of the term linear in $a$ 
becomes opposite.

Eq. (\ref{OE-2}) gives 
the relation between the impact parameter $b$ and  
the inverse of the closest approach $u_0$ as 
$b = u_0^{-1} + M - 2 aM u_0 +O(M^2, a^2)$ 
in the weak field and slow rotation approximations. 
By using this relation, 
$aM$ part of $\phi_{RS}$ in Eq. (\ref{phi-Kerr1}) can be rewritten 
in terms of $b$ as  
\begin{align}
-\frac{2aM}{b^2}\Big[\frac{1}{\sqrt{1-b^2{u_S}^2}}
+\frac{1}{\sqrt{1-b^2{u_R}^2}}\Big] . 
\label{phi-Kerr2}
\end{align}

See Eq. (32) of Ref. \cite{Ishihara2016} 
for $M$ part of $\phi_{RS}$.

\subsection{$\Psi$ parts} 
For the Kerr metric by Eq. (\ref{Kerr}), 
Eq. (\ref{sin}) becomes 
\begin{align}
\sin\Psi
=&\frac{b}{r}\times \cfrac{1-\cfrac{2M}{r}+\cfrac{2aM}{br}}
{\sqrt{1-\cfrac{2M}{r}+\cfrac{a^2}{r^2}}} . 
\end{align}
This is approximated as 
\begin{align}
\sin\Psi 
=&\frac{b}{r} 
\left(1-\frac{M}{r}+\frac{2aM}{br}\right) 
+ 
O\left(\frac{M^2}{r^2}, \frac{a^2}{r^2}, \frac{aM^2}{r^3} \right) . 
\end{align}
By using this, we obtain 
\begin{align}
\Psi_R-\Psi_S
=& \arcsin(bu_R)+\arcsin(bu_S)-\pi 
\notag\\
& -\frac{Mb{u_R}^2}{\sqrt{1-b^2{u_R}^2}}
- \frac{Mb{u_S}^2}{\sqrt{1-b^2{u_S}^2}} 
\notag\\
& +\frac{2aM{u_R}^2}{\sqrt{1-b^2{u_R}^2}} 
+ \frac{2aM{u_S}^2}{\sqrt{1-b^2{u_S}^2}} 
\notag\\
&
+ O\left(M^2 u_R^2, M^2 u_S^2, 
a^2 u_R^2, a^2 u_S^2, 
aM^2 u_R^3, aM^2 u_S^3
\right) . 
\label{Psi-Kerr}
\end{align}

\subsection{Deflection angle of light in Kerr spacetime}
By substituting Eqs. (\ref{phi-Kerr2}) and (\ref{Psi-Kerr}) 
into Eq. (\ref{alpha-axial}),  
the deflection angle of light on the equatorial plane 
in the Kerr spacetime is obtained as
\begin{align}
\alpha_{prog}
=&
\frac{2M}{b}
\left(\sqrt{1-b^2{u_S}^2}+\sqrt{1-b^2{u_R}^2}\right)
\notag\\
&-\frac{2aM}{b^2}
\left(\sqrt{1-b^2{u_R}^2}+\sqrt{1-b^2{u_S}^2}\right)
+ O\left(\frac{M^2}{b^2}\right) , 
\label{alpha+}
\end{align}
where we assume the prograde motion of light. 
For the retrograde case, it is 
\begin{align}
\alpha_{retro}
=&\frac{2M}{b}
\left(\sqrt{1-b^2{u_S}^2}+\sqrt{1-b^2{u_R}^2}\right)
\notag\\
&+\frac{2aM}{b^2}
\left(\sqrt{1-b^2{u_R}^2}+\sqrt{1-b^2{u_S}^2}\right)
+ O\left(\frac{M^2}{b^2}\right) . 
\label{alpha-}
\end{align}
Note that $a^2$ terms at the second order in the deflection angle cancel out. 
See Appendix A for more detail.

For both cases, we take the far limit as $u_R \to 0$ and $u_S \to 0$. 
Then, we obtain 
\begin{align}
\alpha_{\infty\, prog} \to 
&\frac{4M}{b}-\frac{4aM}{b^2} 
+ O\left(\frac{M^2}{b^2}\right) , 
\\
\alpha_{\infty\, retro}\to 
&\frac{4M}{b}+\frac{4aM}{b^2}
+ O\left(\frac{M^2}{b^2}\right) , 
\end{align}
which show that Eqs. (\ref{alpha+}) and (\ref{alpha-}) 
recover the asymptotic deflection angles that are known in literature 
\cite{Ch, ES, Ibanez}.

\subsection{Finite-distance corrections to 
the gravitomagnetic deflection angle of light} 
The above calculations discuss the deflection angle of light 
due to the rotation of the lens (its spin parameter $a$). 
In particular, we do not assume that the receiver and the source 
are located at the infinity. 
The finite-distance correction to the deflection angle of light, 
denoted as $\delta\alpha$, 
is the difference between the asymptotic deflection angle 
$\alpha_{\infty}$ 
and the deflection angle for the finite distance case. 
It is expressed as 
\begin{align}
\delta\alpha \equiv \alpha-\alpha_{\infty} . 
\end{align}
Eqs. (\ref{alpha+}) and (\ref{alpha-})
suggest the magnitude of the finite-distance correction to 
the gravitomagnetic deflection angle by the spin as 
\begin{align}
\left| \delta\alpha_{GM} \right|
\sim& 
O\left(\frac{aM}{r_S^2} + \frac{aM}{r_R^2}\right) 
\notag\\
\sim&O\left(\frac{J}{r_S^2} + \frac{J}{r_R^2}\right) ,  
\label{delta-alpha}
\end{align}
where $J \equiv aM$ is the spin angular momentum of the lens 
and the subscript $GM$ denotes the gravitomagnetic part. 
As usual, we introduce the dimensionless spin parameter as 
$s \equiv a/M$. 
Hence, Eq. (\ref{delta-alpha}) is rewritten as 
\begin{align}
\left| \delta\alpha_{GM} \right| 
\sim O\left( s\left(\frac{M}{r_S}\right)^2 
+ s\left(\frac{M}{r_R}\right)^2 \right) . 
\label{delta-alpha2}
\end{align}
This suggests that $\delta\alpha$ 
is comparable to the second post-Newtonian effect 
(multiplied by the dimensionless spin parameter). 
It is known that the second-order Schwarzschild contribution to $\alpha$ 
is $15\pi M^2/4 b^2$. 
This contribution can be found also by using the present method, 
especially by computing $\phi_{RS}$, where 
we use a relation between $b$ and $r_0$ in $M^2$. 
Please see Appendix A for detailed calculations 
at the second order of $M$ and $a$, 
especially the integrals of $K$ and $\kappa_g$ 
in the present formulation. 
See also the next subsection. 

Note that $\delta\alpha$ at the leading order 
in the approximations does not depend on the impact parameter $b$. 
In fact, $\delta\alpha$ depends much weakly on $b$.

\subsection{Possible astronomical applications}
We discuss possible astronomical applications. 
First, we consider the Sun, 
where we ignore its higher multipole moments. 
The spin angular momentum of the Sun $J_{\odot}$ is 
$\sim 2\times 10^{41} \,\mbox{m}^2\,\mbox{kg}\,\mbox{s}^{-1}$ 
\cite{Sun}. 
Thus, $G J_{\odot} c^{-2} \sim 5 \times 10^5 \,\mbox{m}^2$, 
which implies the dimensionless spin parameter as 
$s_{\odot} \sim 10^{-1}$. 

We assume that an observer at the Earth sees 
the light bending by the solar mass, while 
the source is practically at the asymptotic region. 
If the light ray passes near the solar surface, 
Eq. (\ref{delta-alpha2}) implies that 
the finite-distance correction to this case is 
of the order of 
\begin{align}
\left| \delta\alpha_{GM} \right| 
&\sim 
O \left(\frac{J}{r_R^2}\right) 
\nonumber\\
&\sim 
10^{-12} \mbox{arcsec.} 
\times 
\left(\frac{J}{J_{\odot}}\right) 
\left(\frac{1 \mbox{AU}}{r_R}\right)^2 , 
\label{alpha-Sun}
\end{align}
where $4M_{\odot}/R_{\odot} \sim 1.75 \,\mbox{arcsec.} 
\sim 10^{-5} \,\mbox{rad.}$, 
and $R_{\odot}$ denotes the solar radius. 
This correction is around a pico-arcsecond level and thus 
it is unlikely to be observed with present technology 
\cite{Gaia, JASMINE}. 

\begin{figure}
\includegraphics[width=10cm]{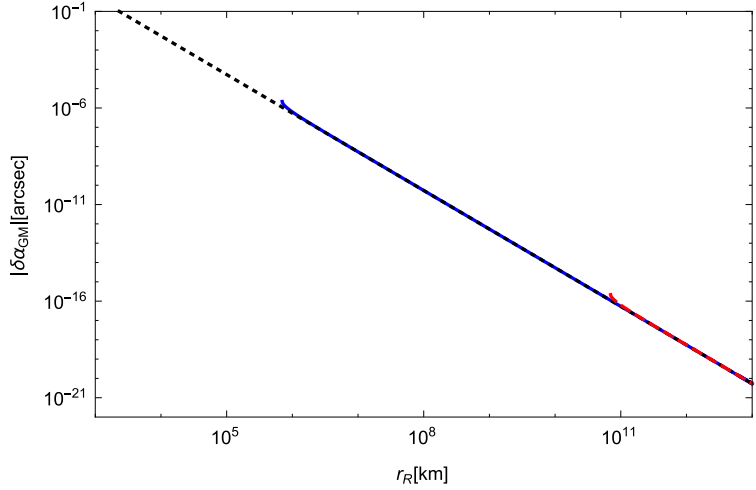}
\caption{ 
$\delta\alpha_{GM}$ for the Sun. 
The vertical axis denotes the finite-distance correction 
to the gravitomagnetic deflection angle of light 
and the horizontal axis denotes the receiver distance $r_R$. 
The solid curve (blue in color) and dashed one (red in color) 
correspond to $b=R_{\odot}$ and $b=10 R_{\odot}$, respectively. 
The dotted line (black in color) denotes the leading term 
of $\delta\alpha_{GM}$ given by Eq. (\ref{delta-alpha}). 
The overlap between these curves suggest that 
the dependence of $\delta\alpha_{GM}$ on the impact parameter $b$ 
is very weak. }
\label{fig-Sun}
\end{figure}

Please see Figure \ref{fig-Sun} for numerical calculations 
of the finite-distance correction due to the receiver location. 
The numerical results are consistent with 
the above order-of-magnitude estimation. 
The figure suggests that the dependence of $\delta\alpha$ 
on the impact parameter $b$ is very weak.

Next, we consider Sgr A$^{\ast}$ at the center of our Galaxy, 
which is expected as 
one of the most plausible candidates for the strong deflection of light. 
In this case, the receiver distance is much larger than 
the impact parameter of light, 
while a source star may be in the central region of our Galaxy. 

For Sgr A$^{\ast}$, Eq. (\ref{delta-alpha}) implies 
\begin{align}
\left| \delta\alpha_{GM} \right| 
&\sim 
s \left( \frac{M}{r_S} \right)^2 
\nonumber\\
&\sim 
10^{-7} \mbox{arcsec.} 
\times 
\left(\frac{s}{0.1}\right) 
\left(\frac{M}{4 \times 10^6 M_{\odot}}\right)^2 
\left(\frac{0.1 \mbox{pc}}{r_S}\right)^2 , 
\label{alpha-Sgr}
\end{align}
where we assume the mass of the central black hole 
as $M \sim 4 \times 10^6 M_{\odot}$. 
This correction around a sub-microarcsecond level 
is unlikely to be measured with present technology. 

\begin{figure}
\includegraphics[width=10cm]{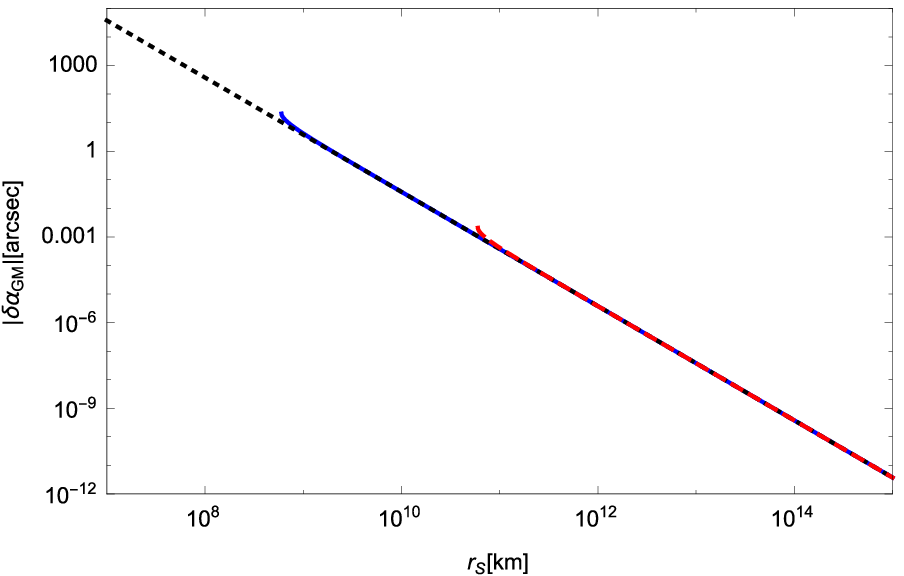}
\caption{ 
$\delta\alpha_{GM}$ 
for the Sgr A$^{\ast}$. 
The vertical axis denotes the finite-distance correction 
to the deflection angle of light 
and the horizontal axis denotes the source distance $r_S$. 
The solid curve (blue in color) and dashed one (red in color) 
correspond to $b=10^2 M$ and $b=10^4 M$, respectively. 
The dotted line (yellow in color) denotes the leading term 
of $\delta\alpha_{GM}$ given by Eq. (\ref{delta-alpha}). 
The overlap between these plots suggest that 
$\delta\alpha_{GM}$ depends faintly on the impact parameter $b$. 
}
\label{fig-Sgr}
\end{figure}

Please see Figure \ref{fig-Sgr} for numerical calculations 
of the finite-distance correction due to the source location. 
The numerical results are consistent with 
the above order-of-magnitude estimation. 
The figure shows that 
the dependence on the impact parameter $b$ is very weak.

\subsection{Consistency of the present formulation}
Before closing this section, let us check the consistency 
of the above formulation. 
The Gaussian curvature is related with 
the 2-dimensional Riemann tensor as 
\cite{Werner2012}
\begin{align}
K
=&\frac{{}^{(3)}R_{r\phi r\phi}}{\gamma} 
\notag\\
=&\frac{1}{\sqrt{\gamma}}
\left[\frac{\partial}{\partial\phi}
\left(\frac{\sqrt{\gamma}}{\gamma_{rr}}{}^{(3)}\Gamma^{\phi}_{~rr}\right)
-\frac{\partial}{\partial r}
\left(\frac{\sqrt{\gamma}}{\gamma_{rr}}{}^{(3)}\Gamma^{\phi}_{~r\phi}\right)
\right] ,
\end{align}
where ${}^{(3)}\Gamma^{i}_{jk}$ and 
${}^{(3)}R_{abcd}$ are associated with $\gamma_{ij}$. 
For the Kerr case, it becomes 
\begin{align}
K
=&
-\sqrt{\frac{A^3}{B(AD+H^2)}}\frac{\partial}{\partial r}
\left[\frac12\sqrt{\frac{A^3}{B(AD+H^2)}}
\frac{\partial}{\partial r}\Big(\frac{AD+H^2}{A^2}\Big)\right] 
\notag\\
=&
-\frac{2M}{r^3}
+ O\left(\frac{M^2}{r^4}, \frac{a^2M}{r^5}
\right) , 
\end{align}
where we use the weak field and slow rotation approximations 
in the last line. 
Note that $K$ has no terms linear in $a$. 
This is because $\gamma_{ij}$ has no terms linear $a$ 
as already mentioned. 
Furthermore, $a^2$ terms cancel out in $K$. 
See Appendix A for more detail.

In order to compute the surface integral of the Gaussian curvature 
in the Gauss-Bonnet theorem, 
we need know the integration domain, 
especially the photon orbit $S \to R$ for the present case. 
By straightforward calculations, 
the iterative solution of Eq. (\ref{OE-2}) for the Kerr case 
in the weak field and slow rotation approximations 
is obtained as 
\begin{align}
u=&\frac{1}{b}\sin\phi+\frac{M}{b^2}(1+\cos^2\phi) 
\notag\\
&-\frac{2aM}{b^3} + O\left(\frac{M^2}{b^3}, \frac{a^2}{b^3}
\right) . 
\end{align}
By using this, 
the surface integral of the Gaussian curvature is computed as 
\begin{align}
-\iint_{{}^{\infty}_{R}\Box^{\infty}_{S}} K dS  
=&
\int^{\infty}_{r_{OE}} 
dr
\int_{\phi_S}^{\phi_R} d\phi 
\frac{2M}{r^2} 
+ O\left(\frac{M^2}{b^2}, \frac{aM^2}{b^3}\right)  
\notag\\
=&
2M \int_{\phi_S}^{\phi_R} d\phi  
\int_{0}^{\frac{1}{b}\sin\phi+\frac{M}{b^2}(1+\cos^2\phi)
-\frac{2aM}{b^3}} du
+ O\left(\frac{M^2}{b^2}, \frac{aM^2}{b^3}\right) 
\notag\\
=&
2M \int_{\phi_S}^{\phi_R} d\phi 
\Big[u\Big]^{\frac{1}{b}\sin\phi+\frac{M}{b^2}(1+\cos^2\phi)
-\frac{2aM}{b^3}}_{u=0}  
+ O\left(\frac{M^2}{b^2}, \frac{aM^2}{b^3}\right) 
\notag\\
=&
\frac{2M}{b} 
\int_{\phi_S}^{\phi_R} d\phi 
\sin\phi  
+ O\left(\frac{M^2}{b^2}, \frac{aM^2}{b^3}\right) 
\notag\\
=&\frac{2M}{b}\Big[\sqrt{1-b^2{u_S}^2}+\sqrt{1-b^2{u_R}^2}\Big] 
+ O\left(\frac{M^2}{b^2}, \frac{aM^2}{b^3}\right) . 
\label{int-K}
\end{align}
It follows that $a^2$ terms do not exist in this calculation.

By combining Eqs. (\ref{int-kappag}) and (\ref{int-K}), 
we obtain 
\begin{align}
-\iint_{{}^{\infty}_{R}\square^{\infty}_{S}} K dS 
- \int_{R}^{S} \kappa_g d\ell 
=& 
\frac{2M}{b}
\left(\sqrt{1-b^2{u_S}^2}+\sqrt{1-b^2{u_R}^2}\right)
\notag\\
&-\frac{2aM}{b^2}
\left(\sqrt{1-b^2{u_R}^2}+\sqrt{1-b^2{u_S}^2}\right) 
\notag\\
&+ O\left(\frac{M^2}{b^2}\right) .  
\label{K-kappag}
\end{align}
This equals to the right-hand side of Eq. (\ref{alpha+}). 
This means that 
the present approach is consistent with the Gauss-Bonnet theorem.

\section{Conclusion}
By using the Gauss-Bonnet theorem in differential geometry, 
we discussed a possible extension of the method of calculating 
the bending angle of light to stationary, axisymmetric 
and asymptotically flat spacetimes. 
We introduced a spatial metric $\gamma_{ij}$ 
to define the bending angle of light, 
which was shown to be coordinate-invariant. 

We considered the light rays on the equatorial plane 
in the axisymmetric spacetime. 
We showed that the geodesic curvature of the photon orbit 
with $\gamma_{ij}$ can be nonzero in gravitomagnetism, 
even though 
the light ray in the four-dimensional spacetime follows 
the null geodesic. 
Finally, we considered Kerr spacetime 
in order to examine how the bending angle of light is computed 
by the present method. 
We made an order-of-magnitude estimate of 
the finite-distance corrections for two possible astronomical cases; 
(1) the Sun and (2) the Sgr A$^{\ast}$. 
The results suggest that 
the finite-distance corrections due to gravitomagnetism 
are unlikely 
to be observed with present technology. 

However, our analysis on possible astronomical observations 
in this paper is limited within the Kerr model. 
It might be interesting to examine the gravitomagnetic 
bending of light by using other axisymmetric spacetimes 
in GR or in a specific theory of modified gravity. 
A further study along this direction is left for future.

\begin{acknowledgments}
We are grateful to Marcus Werner for the stimulating discussions, 
especially for his useful comments 
and his talk on the osculating Riemann approach 
at the seminar in Hirosaki university. 
We would like to thank 
Yuuiti Sendouda, Ryuichi Takahashi, Yuya Nakamura and 
Naoki Tsukamoto 
for the useful conversations. 
This work was supported 
in part by Japan Society for the Promotion of Science 
Grant-in-Aid for Scientific Research, 
No. 26400262 (H.A.), No. 17K05431 (H.A.) and 
in part by by Ministry of Education, Culture, Sports, Science, and Technology,  
No. 15H00772 (H.A.) and No. 17H06359 (H.A.). 
\end{acknowledgments}

\appendix
\section{Detailed calculations at $O(M^2/b^2)$ and $O(a^2/b^2)$}
First, we investigate $K$. 
Up to the second order, it is expanded as 
\begin{align}
K&=\frac{R_{r\phi r\phi}}{\gamma} 
\notag\\
&= -\frac{2M}{r^3}+\frac{3M^2}{r^4}
+O\left(\frac{a^2M}{r^5}\right) , 
\label{K-second}
\end{align}
where $\gamma$ denotes $\det{(\gamma_{ij})}$. 
Note that there are no $a^2$ terms in $K$. 
More interestingly, only the $a^2M$ term among the third order terms 
do exist in $K$. 
By noting that $K$ begins with $O(M)$, what we need 
for the second-order calculations is 
only the linear order in the area element on the equatorial plane. 
This is obtained as 
\begin{align}
dS \equiv& \sqrt{\gamma} drd\phi \notag\\
=& \left[r + 3M + O\left(\frac{M^2}{r}\right) \right] drd\phi , 
\label{dS-second} 
\end{align}
where terms at $O(a)$ and also at $O(a^2)$ do not exist in $dS$. 
This is because all terms including the spin parameter 
cancel out in $\gamma$ for $\theta = \pi/2$ 
and $\gamma$ thus depends only on $M$,  
as can be shown by direct calculations. 

By using Eqs. (\ref{K-second}) and (\ref{dS-second}), 
the surface integration of the Gaussian curvature 
is done as  
\begin{align}
-\iint K dS 
=&\int_{\infty}^{r_{OE}} dr
\int_{\phi_S}^{\phi_R} d\phi \Big(-\frac{2M}{r^3}+\frac{3M^2}{r^4}\Big)
(r+3M) 
+ O\left(\frac{M^3}{b^3}, \frac{aM^2}{b^3}, \frac{a^2M}{b^3}\right)
\notag\\
=&\int_{0}^{\frac{1}{b}\sin\phi+\frac{M}{b^2}(1+\cos^2\phi)} du
\int_{\phi_S}^{\phi_R} d\phi ~(2M+3uM^2) 
+ O\left(\frac{M^3}{b^3}, \frac{aM^2}{b^3}, \frac{a^2M}{b^3}\right)
\notag\\
=&\int_{\phi_S}^{\phi_R}
\Big[2uM+\frac{3u^2}{2}M^2\Big]^{\frac{1}{b}\sin\phi+\frac{M}{b^2}(1+\cos^2\phi)}_{0}
 d\phi 
+ O\left(\frac{M^3}{b^3}, \frac{aM^2}{b^3}, \frac{a^2M}{b^3}\right)
\notag\\
=&\int_{\phi_S}^{\phi_R}
\Big[\frac{2M}{b}\sin\phi+\frac{M^2}{2b^2}(7+\cos^2\phi)\Big] d\phi
+ O\left(\frac{M^3}{b^3}, \frac{aM^2}{b^3}, \frac{a^2M}{b^3}\right)
\notag\\
=&\frac{2M}{b}\Big[\cos\phi\Big]^{\phi_S}_{\phi_R}
+\frac{M^2}{2b^2} \Big[\frac{30\phi+\sin(2\phi)}{4}\Big]^{\phi_R}_{\phi_S}
+ O\left(\frac{M^3}{b^3}, \frac{aM^2}{b^3}, \frac{a^2M}{b^3}\right)
\notag\\
=&\frac{2M}{b}\Big[\sqrt{1-b^2{u_S}^2}+\sqrt{1-b^2{u_R}^2} \Big] 
\notag\\
&+\frac{2M^2}{b}\Big[\frac{u_S(2-b^2{u_S}^2)}{\sqrt{1-b^2{u_S}^2}}
+\frac{u_R(2-b^2{u_R}^2)}{\sqrt{1-b^2{u_R}^2}}\Big]
\notag\\
&+\frac{15 M^2}{4b^2}[\pi-\arcsin(bu_S)-\arcsin(bu_R)]
\notag\\
&-\frac{M^2}{4b^2}[bu_S\sqrt{1-b^2{u_S}^2}+bu_R\sqrt{1-b^2{u_R}^2}] 
+ O\left(\frac{M^3}{b^3}, \frac{aM^2}{b^3}, \frac{a^2M}{b^3}\right)
\notag\\
=&\frac{2M}{b}\Big[\sqrt{1-b^2{u_S}^2}+\sqrt{1-b^2{u_R}^2}\Big] 
\notag\\
&+\frac{15 M^2}{4b^2}[\pi-\arcsin(bu_S)-\arcsin(bu_R)]
\notag\\
&+\frac{M^2}{4b^2}[\frac{bu_S(15-7b^2{u_S}^2)}{\sqrt{1-b^2{u_S}^2}}
+\frac{bu_R(15-7b^2{u_R}^2)}{\sqrt{1-b^2{u_R}^2}}] 
+ O\left(\frac{M^3}{b^3}, \frac{aM^2}{b^3}, \frac{a^2M}{b^3}\right) , 
\label{int-K-second}
\end{align}
where we use, in the second line, 
an iterative solution for the orbit equation 
by Eq. (\ref{OE}) for the Kerr spacetime. 

Next, we study the geodesic curvature. 
On the equatorial plane, we obtain 
\begin{align}
\kappa_g=&-\frac{1}{\sqrt{\cfrac{\Sigma^2}{\Delta(\Sigma-2Mr)}
\left(r^2+a^2+\cfrac{2a^2Mr\sin^2\theta}{\Sigma}\right)
\cfrac{\Sigma\sin^2\theta}{(\Sigma-2Mr)}}}\beta_{\phi,r} \notag \\
=&-\frac{2aM}{r^3} +O\left( \frac{aM^2}{r^3}\right) , 
\label{kappag-second}
\end{align}
where $a^2$ terms do not exist. 
From this, we obtain 
\begin{align}
\int_{c_p}\kappa_gd\ell
=&-\int^{R}_{S} d\ell \left[ \frac{2aM}{r^2} 
+ O\left( \frac{aM^2}{r^3} \right) \right] \notag\\
=&-\frac{2aM}{b^2}\int^{\phi_R}_{~\phi_S} \cos\vartheta d\vartheta 
+ O\left( \frac{aM^2}{b^3} \right)
\notag\\
=&-\frac{2aM}{b^2}[\sin\phi_R-\sin\phi_S] 
+ O\left( \frac{aM^2}{b^3} \right)
\notag\\
=&\frac{2aM}{b^2}[\sqrt{1-b^2{u_R}^2}+\sqrt{1-b^2{u_S}^2}] 
+ O\left( \frac{aM^2}{b^3} \right) , 
\label{int-kappag-second}
\end{align}
where we use 
$\sin\phi_S=\sqrt{{r_S}^2-b^2}/r_S + O(M/r_S)$ 
and  
$\sin\phi_R=-\sqrt{{r_R}^2-b^2}/r_R + O(M/r_R)$. 

By combining Eqs. (\ref{int-K-second}) and (\ref{int-kappag-second}), 
we obtain 
\begin{align}
\alpha \equiv& -\iint_{{}^{\infty}_{R}\square^{\infty}_{S}} K dS 
-\int^{S}_{R}\kappa_gd\ell \notag\\
=&\frac{2M}{b}\Big[\sqrt{1-b^2{u_S}^2}+\sqrt{1-b^2{u_R}^2}\Big] 
\notag\\
&+\frac{15 M^2}{4b^2}\left[\pi-\arcsin(bu_S)-\arcsin(bu_R)\right]
\notag\\
&+\frac{M^2}{4b^2}\left[\frac{bu_S(15-7b^2{u_S}^2)}{\sqrt{1-b^2{u_S}^2}}
+\frac{bu_R(15-7b^2{u_R}^2)}{\sqrt{1-b^2{u_R}^2}}\right] \notag\\
&-\frac{2aM}{b^2}\left[\sqrt{1-b^2{u_R}^2}+\sqrt{1-b^2{u_S}^2}\right] 
+ O\left(\frac{M^3}{b^3}, \frac{aM^2}{b^3}, \frac{a^2M}{b^3}\right) . 
\label{alpha-2nd}
\end{align}
Note that $a^2$ terms and $a^3$ ones do not appear in $\alpha$ 
for the finite distance situation as well as 
in the infinite distance limit. 
If we assume the infinite distance limit $u_R, u_S \to 0$, 
Eq. (\ref{alpha-2nd}) becomes 
\begin{align}
\alpha \rightarrow \frac{4M}{b}+\frac{15\pi M^2}{4b^2}-\frac{4aM}{b^2} . 
\end{align}
This agrees with the previous results, 
especially the numerical coefficients at the order of $M^2$ and $aM$.

\newpage

\begin{thebibliography}{99}
\bibitem{Kerr-bending}
In a conventional approach, 
precise analytic treatments of the deflection angle of light 
were done 
in the equatorial plane of a Kerr black hole \cite{IH} 
and for generic photon orbits 
in terms of the generalized hypergeometric functions 
of Appell and Lauricella \cite{Kraniotis}, 
though their calculations are entirely based on an assumption that 
both the source and the receiver are located at the null infinity.
\bibitem{Metric}
In this paper, we use the polar coordinates. 
In the cylindrical coordinates, the line element is known 
as the Weyl-Lewis-Papapetrou form \cite{Lewis,LR,Papapetrou}.
\bibitem{comment-gamma}
In this section and after, 
we do not call $\gamma_{ij}$ the optical metric, 
because the photon orbit is not always 
a geodesic in a space defined by $\gamma_{ij}$. 
See Eq. (\ref{EL}).
\end{thebibliography}
\end{document}